\documentstyle[11pt,newpasp,twoside,epsf]{article}
\def\edcomment#1{\iffalse\marginpar{\raggedright\sl#1\/}\else\relax\fi}
\reversemarginpar

\begin{document}

\title{Clustered vs.\ Isolated Star Formation}

\author{Mordecai-Mark Mac Low}
\affil{Department of Astrophysics, American Museum of Natural History, 79th Street at Central Park West, New York, NY, 10024-5192, USA}

\begin{abstract}
I argue that star formation is controlled by supersonic turbulence,
drawing for support on a number of 3D hydrodynamical and MHD
simulations as well as theoretical arguments.  Clustered star
formation appears to be a natural result of a lack of turbulent
support, while isolated star formation is a signpost of global
turbulent support.
\end{abstract}

\section{Introduction}
The big question in star formation is not how to form stars, but how
to prevent stars from forming.  In the Milky Way, slow, continuous
star formation is seen, despite the presence of a great deal of gas.
The free-fall time for gas is $t_{\rm ff} = (3\pi/32 G \rho)^{1/2} = (1.2
\times 10^7 \mbox{ yr})(n/10 \mbox{ cm}^{-3})^{-1/2}$, where $n$ is
the number density of the cloud, and I assume the mean molecular mass
$\mu = \rho/n = 3.32 \times 10^{-24}$ g.  Yet, much of this gas must
have been around for times of order the Galaxy lifetime of $10^{10}$
yr.  How can we explain this?  Galaxies show a wide variety of star
formation rates, ranging from low surface brightness galaxies with
plenty of gas and virtually no star formation, to starbursts with star
formation rates sufficient to consume their gas in a small fraction of
a Hubble time if sustained.  How can we explain the variation between
galaxies?

One way out of this problem is for molecular clouds to last far longer
than a free-fall time, as was suggested by Blitz \& Shu (1980), who
argued on the grounds of star formation rate, ages of stars associated
with clouds, and positions of clouds behind spiral shocks, for cloud
lifetimes of order 30~Myr.  Recently, though, much shorter lifetimes
for molecular clouds of only 5--10~Myr have been suggested by
Ballesteros-Paredes, Hartmann, \& V\'azquez-Semadeni (1999), based on
the lack of observed post-T Tauri stars associated with molecular
clouds, and the observation that density enhancements in interstellar
medium (ISM) simulations are created and destroyed quickly.

Molecular clouds are observed to have linewidths much broader than
thermal, created by hypersonic random motions in the clouds.  Such
random motions produce strong shocks, which ought to dissipate the
energy in roughly a crossing time.  The question of how the observed
motions can be maintained appears intimately intertwined with the
question of how the clouds avoid collapse on a free-fall time.

In what might be called the standard theory of star formation,
magnetic fields are invoked to answer both of these questions.  If
fields are strong enough, they can magnetostatically support clouds
against collapse.  The star formation rate would then be determined by
the rate of ambipolar drift of neutral gas past ions tied to the
magnetic field towards the centers of self-gravitating cores
(Mouschovias 1977, Shu 1977).  Furthermore, if the fields are strong
enough that the Alfv\'en speed $v_A$ reaches the rms velocity $v$,
then strong shocks will be converted to MHD waves.  As linear Alfv\'en
waves are lossless, it was thought that motions remaining from the
initial formation of the clouds might be enough to explain the
observation (Arons \& Max 1975).

In this review I will explain why both of these ideas now appear
questionable.  I will call on computations performed with two
different methods: Eulerian hydrodynamics and MHD on a grid, using the
code ZEUS-3D (Stone \& Norman 1992a, b; Clarke 1994; Hawley \& Stone
1995; available from the Laboratory for Computational Astrophysics at
http://zeus.ncsa.uiuc.edu/lca\_home\_page.html), and Lagrangian
hydrodynamics using a smoothed particle hydrodynamics (SPH) code
derived from that described by Benz (1990) and Monaghan (1992),
running on special purpose GRAPE processors (Ebisuzaki et al.\ 1993;
Steinmetz 1996), and incorporating sink particles (Bate, Bonnell, \&
Price 1995).

\section{Energy Dissipation}

Does trans-Alfv\'enic turbulence decay at a rate substantially slower
than hypersonic turbulence?  Mac Low et al.\ (1998) tested this idea
by directly computing the decay of 3D hypersonic turbulence with and
without an initially uniform magnetic field in a domain with periodic
boundary conditions.  They found that the kinetic energy of
unmagnetized turbulence decayed following a law $\dot{E}_{kin} \propto
t^{-\eta}$, with $\eta = 1$.  This result was supported by comparison
between ZEUS and SPH results using resolution studies over models
ranging from $32^3$ to $256^3$ zones and $19^3$ to $70^3$ particles,
respectively.  Adding weak magnetic fields, with initial sound speed
$c_s = v_A$, reduced the decay rate only slightly, to $\eta = 0.91$.
Even a strong magnetic field, with initial $v_A = v$, decayed with
$\eta = 0.87$.  While the difference in decay rates is of interest to
theorists seeking to understand the detailed behavior of MHD
turbulence, it is clearly insufficient to explain the observed motions
in molecular clouds as coming from their initial conditions.  It
appears that the interaction of a full set of MHD waves in 3D
transfers sufficient power to waves with wavelengths of order the
dissipation scale, whatever it may be, to dissipate energy quickly.

To quantify the decay rate of hypersonic and compressible MHD
turbulence, Mac Low (1999) used a uniform, fixed pattern of Gaussian
perturbations to the velocity field with typical wavenumber $k$ to
drive the turbulence with a fixed energy input rate.  The squares in
Fig.~1 show that the measured energy dissipation rates
for hypersonic unmagnetized turbulence follow the pattern expected
from dimensional analysis, with the actual rate given by
\begin{equation}
 \dot{E}_{kin} = (0.21/\pi) m k v^3, \label{edot}
\end{equation}
where $m$ is the mass in the cube. The triangles show results from
magnetized models with varying field strengths.  Weak field models
appear to diverge more from the hydrodynamical result as the fields
are tangled than strong-field models, where the flow is organized by
the field into a roughly 1D flow along the field lines.

\begin{figure}[tp]
\plottwo{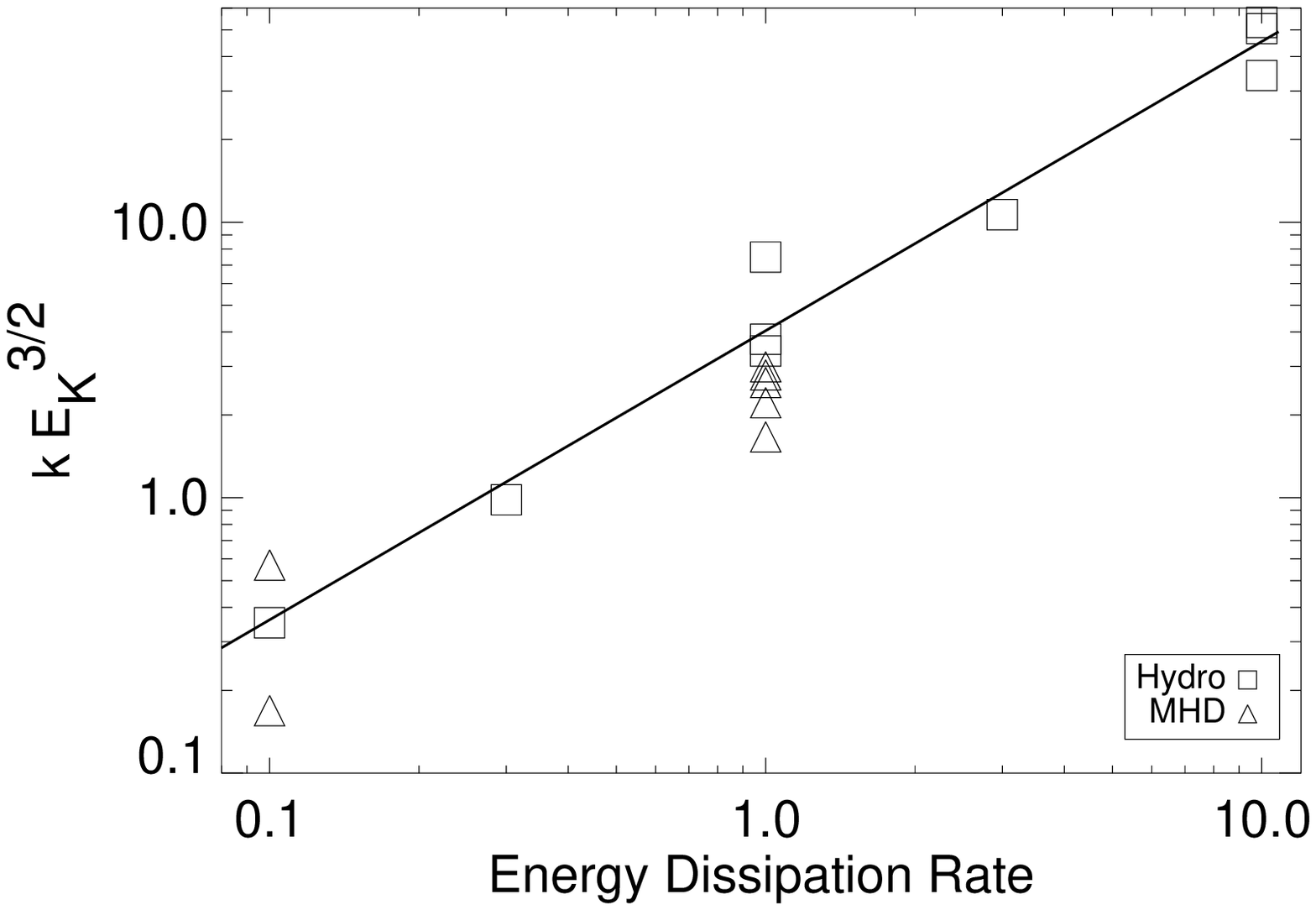}{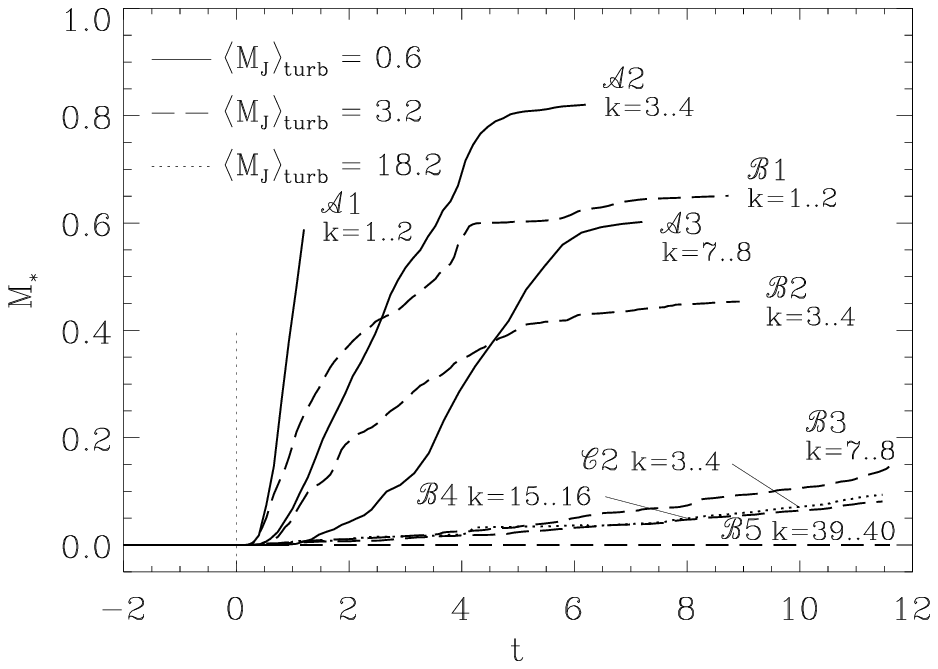}
\caption{\footnotesize ({\bf left}) Kinetic energy $E_K$ needed in
uniformly driven turbulence to maintain varying energy dissipation
rates with dimensionless driving wavenumbers $k$.  The line shows the
relation given in eq.~(1).  From Mac Low (1999).  ({\bf right})
Fraction of mass in condensed objects against time in free-fall times
for different SPH models with driving wavenumbers $k$ and turbulent
Jeans masses given.  The mass of the region in each case is unity;
larger turbulent Jeans masses imply stronger driving. From Klessen,
Heitsch, \& Mac Low (2000).
\label{figdrive}}
\end{figure}

Knowing the decay rate allows calculation of the formal decay time
$t_d = E_{kin}/\dot{E}_{kin}$ in terms of the free-fall time $t_{\rm ff} =
\lambda_J/c_s$, where $\lambda_J = c_s (\pi/G)^{1/2}$ is the Jeans
wavelength. Taking their ratio, Mac Low (1999) finds that 
\begin{equation}
\frac{t_d}{t_{\rm ff}} = 1.2 \frac{\lambda_d}{\lambda_J}
\frac{\pi}{M_{rms}} \ll 1,
\end{equation}
where $\lambda_d = 2\pi/k$ is the driving wavelength, and $M_{rms} =
v/c_s$ is the rms Mach number.  The driving wavelength needs to be
shorter than the Jeans wavelength to provide turbulent support against
collapse, as suggested by Bonazzola et al.\ (1987, 1992), while
the typical observed Mach number is substantially higher than $\pi$,
so turbulence capable of supporting molecular clouds decays in
substantially less than a free-fall time.  Thus, the observed motions
cannot come from initial conditions unless the clouds are very young,
or the motions are very long wavelength.

\section{Turbulent Support}

Magnetohydrostatic support of molecular cloud cores balanced by
ambipolar diffusion allowing collapse appears attractive because it
can extend the star formation timescale by more than an order of
magnitude.  Strong magnetic fields are indeed observed in very dense
regions in molecular clouds with $n > 10^6$~cm$^{-3}$, as can be
deduced from water maser measurements (e.g.\ Elitzur, Hollenbach, \&
McKee 1992).  

However, recent measurements of magnetic field strengths in lower
density regions using OH Zeeman measurements have gradually been
leading to the conclusion that the field is roughly a factor of two
lower than required for magnetostatic support (Crutcher 1999),
although the error bars are still of the same size as the measurement.
Another piece of evidence against magnetostatic support is the
column density contrast of observed cores, which is far higher than
would be expected for subcritical, magnetically supported cores
(Nakano 1998).  Finally, the ambipolar diffusion picture would predict
that roughly one core in seven had a condensed protostar in its
center, while ISO observations reveal as many as one in two
(Ward-Thompson, Motte, \& Andre 1999).

Supersonic turbulence offers an alternative support mechanism against
collapse, as reviewed by Scalo (1985).  Analytic work by L\'eorat et
al.\ (1990) and Bonazzola et al.\ (1988, 1992) suggested that it would
indeed be effective in preventing collapse, but only if the driving
wavelength $\lambda_d < \lambda_J$.  Numerical models including
self-gravity were used to directly test how effective supersonic
turbulence is at supporting a region against collapse.

Quantification of star formation rates from our simulations is
difficult, as sufficient resolution to follow the collapse and
fragmentation of protostellar cores, as specified by Truelove et
al. (1997) for grid codes and Bate \& Burkert (1997) for SPH, will
require adaptive mesh refinement techniques.  Instead, we bracket the
true behavior with ZEUS and sink particles in the SPH code.  Collapsed
regions in ZEUS cannot collapse past the grid scale, so remain far too
large, and too easily destroyed by passing shocks.  On the other hand,
mass swallowed by sink particles is never given up, so they give
an upper limit to the amount of mass going into the cores, and
ultimately into stars.

A region of gas at rest, or one initialized with turbulent motions
that are allowed to decay both collapse efficiently (Klessen, Burkert,
\& Bate 1998; Klessen \& Burkert 2000).  If we add driving, we get the
results shown in the right panel of Figure~1, which shows the fraction
of mass going into condensed objects for a number of SPH runs with
different driving parameters.  All of these runs are formally
supported against collapse by the turbulent Jeans criterion,
$<M_J>_{\rm turb} = v^3 (3\pi/32 G \rho)^{1/2} > m$, where $m$ is the
mass of the region.  Nevertheless, as much as 80\% of the mass ends up
in collapsed cores, depending on the driving parameters.  Stronger
driving (larger turbulent Jeans mass) and shorter wavelength driving
both inhibit collapse.

The collapse rate can be decreased to values consistent with those
observed of a few percent over tens of free-fall times with strong
enough driving.  To completely prevent collapse, however, requires
driving not just strong enough and at short enough wavelengths to
support the average density, but rather values a factor of the Mach
number $M$ stronger and shorter.  Not only must the whole region be
supported, but the density enhancements caused by isothermal shocks
must also be supported.  Isothermal shocks cause compressions
proportional to $M^2$, so $<M_J>_{\rm turb} \propto \rho^{-1/2}$ must
be increased by $M$ to support them, and correspondingly for the
driving wavelength. Magnetic fields do not qualitatively change this
conclusion (Heitsch, Mac Low, \& Klessen 2001), although they do
reduce the rate of collapse for any particular level of driving.  {\em
Isolated regions of collapse are thus an observational sign of overall
turbulent support.}

Furthermore, the distribution of resulting condensed objects depends
on the properties of the driving as shown in Figure~3.  Strong,
short-wavelength driving results in condensed cores distributed evenly
across the region, reminiscent of low-rate star formation as is seen,
for example, in Taurus.  Long-wavelength driving, or the absence of
driving, on the other hand, leads to clustered core formation, with
most of the mass ending up in a rather small region of the total
volume.  This reproduces the clustering seen not only in regions of
massive star formation like Orion, but also in starburst regions in
nearby and distant galaxies.
\begin{figure}[tp]
\plotfiddle{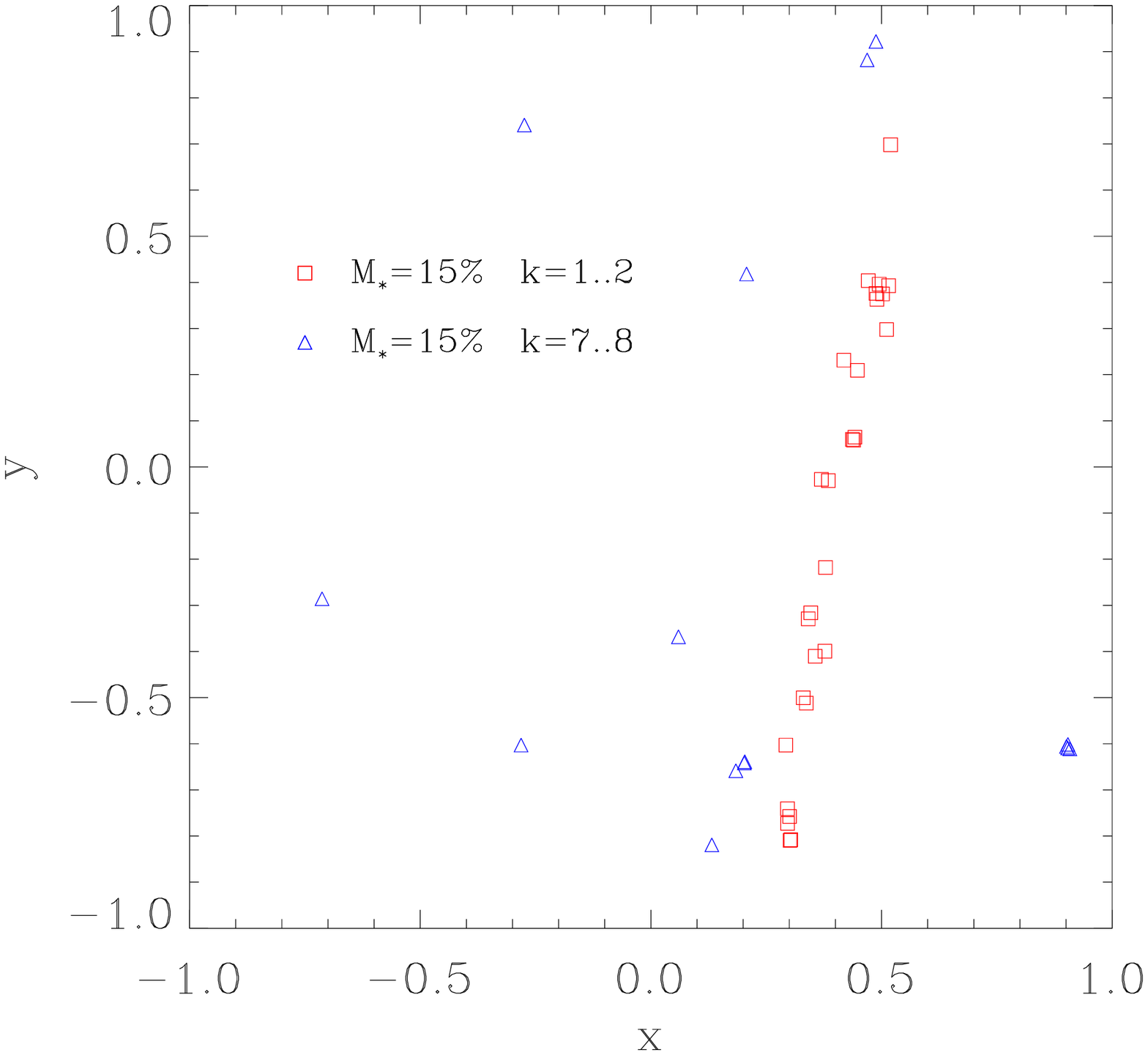}{1in}{0}{25}{25}{-150}{-68}
\plotfiddle{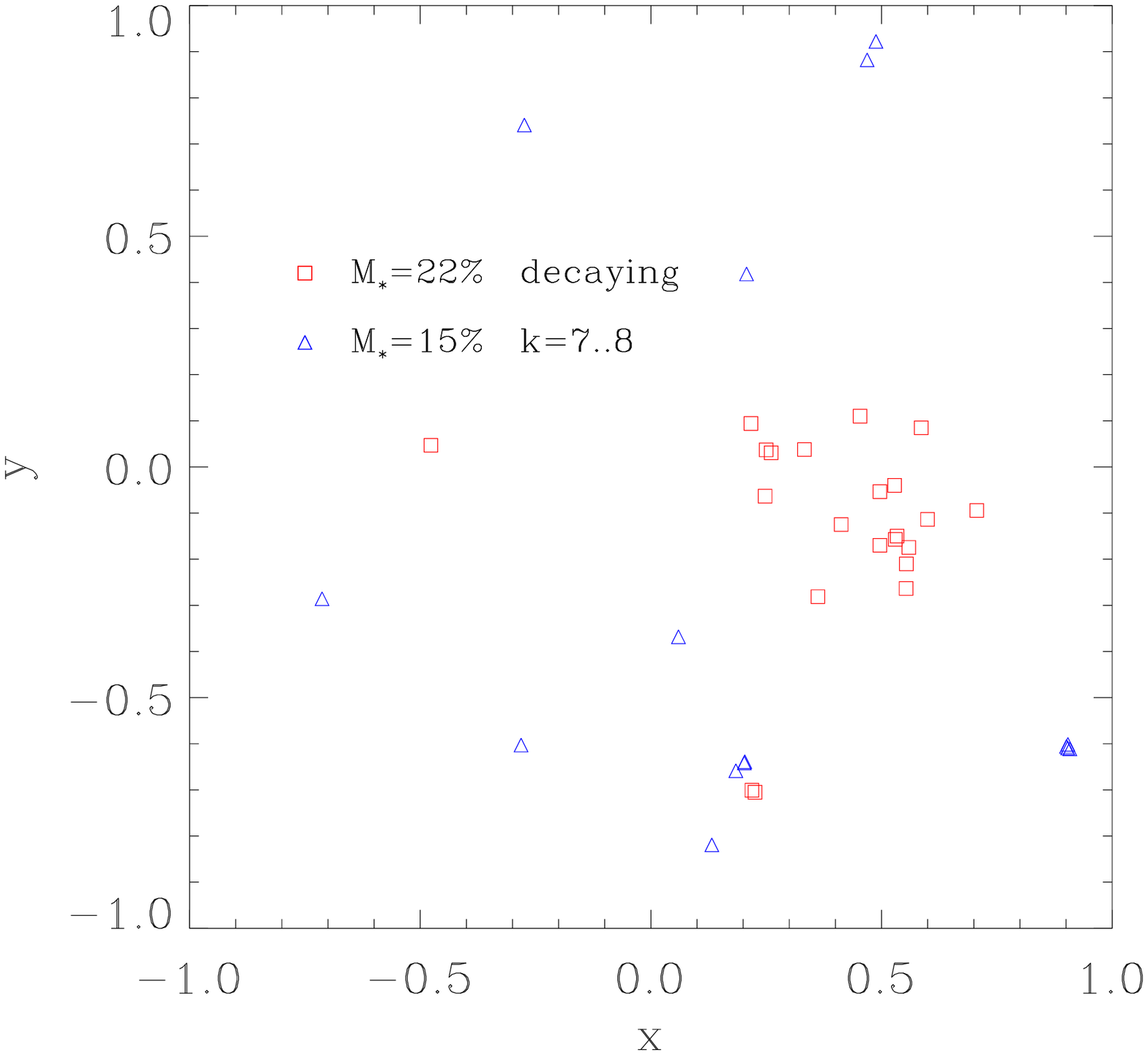}{0in}{0}{25}{25}{0}{-43}
\caption{\footnotesize Projected locations of sink particles in SPH
models. (left) Comparison between models driven at large scales
(squares) and small scales (triangles), showing that the driving
wavelength partially determines clustering.  (right) Comparison
between driven and decaying models showing the difference between
local collapse in a globally supported region (triangles) and global
collapse in an unsupported region (squares).  (Klessen et al.\ 2000)}
\end{figure}

\section{Driving Mechanisms}

Both support against gravity and maintenance of observed motions
appear to depend on continued driving of the turbulence, as I have
described.  What then is the energy source for this driving?

Motions coming from gravitational collapse have often been suggested
but fail due to the quick decay of the turbulence as described above.
If the turbulence decays in less than a free-fall time, then it cannot
delay collapse for substantially longer than a free-fall time (Klessen
\& Burkert 2000).

Protostellar jets and outflows are another popular suspect for the
energy source of the observed turbulence.  They are indeed quite
energetic, but they deposit most of their energy into low density gas,
as is shown by the observation of multi-parsec long jets extending
completely out of molecular clouds (Bally \& Devine 1994).
Furthermore, the observed motions show increasing power on scales all
the way up to and perhaps beyond the largest scale of molecular cloud
complexes (Ossenkopf \& Mac Low 2001).  It is hard to see how such
large scales could be driven by protostars embedded in the clouds.

Another energy source that has long been considered is shear from
galactic rotation.  Recent work by Sellwood \& Balbus (1999) has shown
that magnetorotational instabilities (Balbus \& Hawley 1991, 1998)
could couple the large-scale motions to small scales efficiently.  For
parameters appropriate to the far outer H~{\sc i} disk of the Milky
Way, they derive a resulting velocity dispersion of 6~km~s$^{-1}$,
close to that observed.  This instability may provide a base value for the
velocity dispersion below which no galaxy will fall.  If that is
sufficient to prevent collapse, little or no star formation will
occur, producing something like a low surface brightness galaxy with
large amounts of H~{\sc i} and few stars.

In active star-forming galaxies, however, clustered and field supernova
explosions, predominantly from B~stars no longer associated with their
parent gas, appear likely to dominate the driving, raising the
velocity dispersion to the 10--15 km~s$^{-1}$ observed in star-forming
portions of galaxies (see work cited in Mac Low 2000 for example).
This provides a large-scale self-regulation mechanism for star
formation in disks with sufficient gas density to collapse despite the
velocity dispersion produced by the magnetorotational instability.  As
star formation increases in such galaxies, the number of OB stars
increases, ultimately increasing the supernova rate and thus the
velocity dispersion, which will restrain further star formation.  

Supernova driving not only determines the velocity dispersion, but may
actually form molecular clouds by sweeping gas up in a turbulent
flow.  A snapshot of such a flow is shown in Figure~4.  The densest
regions are formed by shock-wave interactions (cooling is included,
but not self-gravity) on a dynamical timescale, and disperse on the
same short timescale.  The domain shown is 1~kpc$^2$, giving dynamical
timescales of a few million years, as estimated by Ballesteros-Paredes
et al.\ (1999) for the lifetime of molecular clouds.  
\begin{figure}[tp]
\plotone{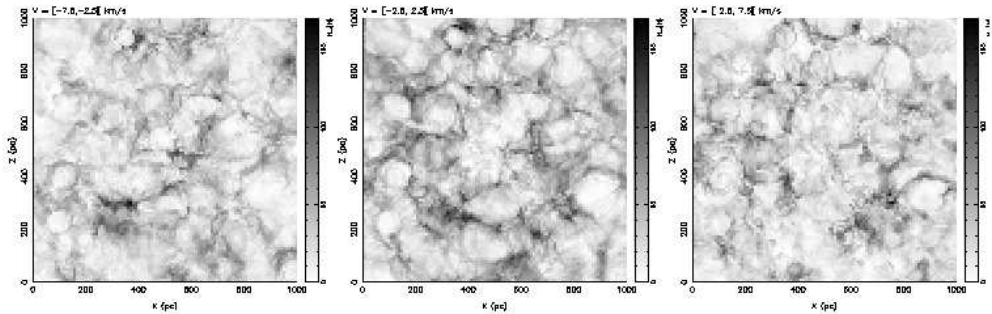}
\caption{\footnotesize Simulated observations of a supernova-dominated
ISM from the model described by Avillez (2000).  These are images of
column density in three velocity intervals separated by 5~km~s$^{-1}$,
viewed from above, in a $1\times1\times10$ kpc simulation domain. The
densest regions are likely to form molecular gas.  These slices may be
directly compared to the observations of the LMC by Kim, Dopita, \&
Stavely-Smith (1999). \label{figavillez}}
\end{figure}

\section{Final Thoughts}
Support by driven supersonic turbulence balanced against self-gravity
appears able to explain a number of things, including the timescales
for star formation, the observed supersonic motions in molecular
clouds, their filamentary, clumpy morphology, the different modes of
star formation observed, and ultimately the difference between steady,
low-efficiency star formation and starburst behavior.  It is
consistent with observed magnetic fields, although they do not play a
central role.  Open questions in this picture are then: 1) Can we derive a
quantitative star formation rate given the velocity dispersion and
local density of the gas?  2) What determines global behavior such as
the Schmidt law for star formation?  3) Are supernovae the primary
mechanism driving the turbulence?

\acknowledgements {\footnotesize I thank the organizers for their kind
invitation and support of my attendance, and the US National Science
Foundation for partial support of this work by a CAREER fellowship
AST99-85392. I thank my collaborators for many discussions of the
material presented here.  Computations shown here have been performed
in part at the National Center for Supercomputing Applications and the
Rechenzentrum Garching of the MPG.  This work has made use of the NASA
Astrophysical Data System Abstract Service.}

\end{document}